\documentclass[aps,preprint,nofootinbib,superscriptaddress, floatfix,groupedaddress,longbibliography]{revtex4-1}
\usepackage[english]{babel}
\usepackage[utf8]{inputenc}
\usepackage{mathtools}
\usepackage{braket}
\usepackage{amsfonts}
\usepackage{lineno}
\usepackage{pdfpages}
\usepackage{stfloats}
\usepackage{xcolor}
\usepackage[symbol]{footmisc}
\makeatletter
\AtBeginDocument{\let\LS@rot\@undefined}
\makeatother

\usepackage[justification=raggedright,singlelinecheck=false]{caption,subcaption}
\usepackage{siunitx}

\usepackage{scalerel}
\usepackage{tikz}
\usetikzlibrary{svg.path}
\definecolor{orcidlogocol}{HTML}{A6CE39}
\tikzset{
  orcidlogo/.pic={
    \fill[orcidlogocol] svg{M256,128c0,70.7-57.3,128-128,128C57.3,256,0,198.7,0,128C0,57.3,57.3,0,128,0C198.7,0,256,57.3,256,128z};
    \fill[white] svg{M86.3,186.2H70.9V79.1h15.4v48.4V186.2z}
                 svg{M108.9,79.1h41.6c39.6,0,57,28.3,57,53.6c0,27.5-21.5,53.6-56.8,53.6h-41.8V79.1z M124.3,172.4h24.5c34.9,0,42.9-26.5,42.9-39.7c0-21.5-13.7-39.7-43.7-39.7h-23.7V172.4z}
                 svg{M88.7,56.8c0,5.5-4.5,10.1-10.1,10.1c-5.6,0-10.1-4.6-10.1-10.1c0-5.6,4.5-10.1,10.1-10.1C84.2,46.7,88.7,51.3,88.7,56.8z};}}
\newcommand\orcidicon[1]{\href{https://orcid.org/#1}{\mbox{\scalerel*{
\begin{tikzpicture}[yscale=-1,transform shape]
\pic{orcidlogo};
\end{tikzpicture}
}{|}}}}
\usepackage[colorlinks,citecolor=red,urlcolor=blue,bookmarks=false,hypertexnames=true]{hyperref} 
\usepackage[capitalise]{cleveref}

\newcommand{\kwv}{\mathbf{k}}
\newcommand{\kv}{$\rm kV \, cm^{-1}$}

\newcommand{\vcm}{$\rm V \, cm^{-1}$}

\begin{document}

\title{Hot electron diffusion, microwave noise, and piezoresistivity in Si from first principles}

\author{Benjamin Hatanp{\"a}{\"a} \orcidicon{0000-0002-8441-0183}}
\affiliation{Division of Engineering and Applied Science, California Institute of Technology, Pasadena, CA, USA}

\author{Austin J. Minnich \orcidicon{0000-0002-9671-9540}}
\thanks{Corresponding author: \href{mailto:aminnich@caltech.edu}{aminnich@caltech.edu}}

\affiliation{Division of Engineering and Applied Science, California Institute of Technology, Pasadena, CA, USA}

\date{\today} 

\begin{abstract}

Ab-initio calculations of charge transport properties in materials without adjustable parameters have provided microscopic insights into electron-phonon interactions which govern charge transport properties. Other transport properties such as the diffusion coefficient provide additional microscopic information and are readily accessible experimentally, but few ab-initio calculations of these properties have been performed. Here, we report first-principles calculations of the hot electron diffusion coefficient in Si and its dependence on electric field over temperatures from 77 -- 300 K. While qualitative agreement in trends such as anisotropy at high electric fields is obtained, the quantitative agreement that is routinely achieved for low-field mobility is lacking. We examine whether the discrepancy can be attributed to an inaccurate description of f-type intervalley scattering by computing the microwave-frequency noise spectrum and piezoresistivity. These calculations indicate that any error in the strength of f-type scattering is insufficient to explain the diffusion coefficient discrepancies. Our findings suggest that the measured diffusion coefficient is influenced by factors such as space charge effects which are not included in ab-initio calculations, impacting the interpretation of this property in terms of charge transport processes.

\end{abstract}

\maketitle

\noindent

\section{Introduction}

Ab-initio calculations of linear transport coefficients such as the electrical mobility of materials without adjustable parameters are now routine \cite{Li_2015,Fiorentini_2016,Ponce_2018,Zhou_2016,Liu_2017}. The approach is based on density functional theory (DFT) and density functional perturbation theory (DFPT) to compute the electronic structure, phonons, and electron-phonon matrix elements, followed by Wannier interpolation to the fine grids needed for transport calculations \cite{Bernardi_2016,Giustino_2017}. Transport properties are obtained by solving the Boltzmann equation with the collision matrix computed from the ab-initio inputs. The accuracy of these calculations for low-field mobility has been established for many materials, including Si \cite{Fiorentini_2016,Li_2015,Ponce_2021}, GaAs \cite{Zhou_2016,Ponce_2021,Liu_2017}, and others \cite{Ponce_2020}.

While ab-initio calculations of low-field mobility are relatively mature, transport properties beyond the low-field regime and diffusion coefficients have historically been evaluated using Monte Carlo methods \cite{Jacoboni_1975,Brunetti_1981,Jacoboni_1983,Pop_2004}. These models utilized various approximations, such as dispersionless optical phonons, Debye acoustic phonons, and model bandstructures. More recently, the Monte Carlo method in n-Si has been extended to device simulation \cite{Aksamija_2006} and full-band studies \cite{Fischetti_1991,Fischer_2000, Nguyen_2003,Fischetti_2019}.

Computations of high-field transport properties using ab-initio methods have only recently been reported \cite{Choi_2021,Maliyov_2021,Cheng_2022,Sun_2023,Catherall_2023}. In GaAs, the drift velocity characteristics up to several \kv~have been computed and have provided evidence for the role of two-phonon scattering \cite{Cheng_2022}. The warm electron tensor of Si has also been computed and directly compared to experiment \cite{Hatanpaa_2023}. Another experimentally accessible property is the power spectral density (PSD) of current fluctuations, which in the low-frequency limit is proportional to the diffusion coefficient \cite{Hartnagel_2001}. As this quantity is more sensitive to certain details of the band structure and scattering rates than mean charge transport properties  \cite{Hartnagel_2001}, it provides a stricter test for ab-initio methods compared to the low-field mobility. An ab-initio formalism to calculate fluctuational properties has been developed recently \cite{Choi_2021,Cheng_2022,Catherall_2023}.

Certain noise phenomena such as intervalley noise, where carrier number fluctuations between valleys cause current fluctuations if the drift velocities in different valleys differ, can only be observed in a multi-valley semiconductor such as n-Si \cite{Price_1960, Hartnagel_2001}. The PSD of hot electrons in Si has been experimentally investigated at a range of frequencies, temperature, and electric field strengths. Measurements of the electron diffusion coefficient (proportional to low-frequency PSD) at room temperature along the [111] direction indicated a pronounced decrease with increasing electric field \cite{Canali_1975_diffusion}. A subsequent study at lower temperatures showed an initial increase of the diffusion coefficient with increasing field, followed by the decrease seen at higher temperatures \cite{Brunetti_1981}. A clear anisotropy in the diffusion coefficient was observed between the [100] and [111] directions for high electric fields ($\gtrsim 2$ \kv) across 77 -- 300 K, \cite{Brunetti_1981} despite the cubic symmetry of Si, and was attributed to intervalley noise. The frequency-dependence of the PSD at low temperatures $\sim 80$ K was also examined, showing how thermal, convective, and intervalley noise contribute at different frequencies for these two directions \cite{Bareikis_1982}. However, whether ab-initio methods can accurately account for these observed transport properties has not yet been determined.

Here, we report first-principles calculations of the hot electron diffusion coefficient in Si. We find that although some qualitative features of the diffusion coefficient are correctly predicted, such as an anisotropy at high electric fields, quantitative agreement is in general poor. To identify the origin of the discrepancies, we computed the microwave-frequency PSD and piezoresistivity. The computed properties lack the qualitative disagreements with experiment found for the diffusion coefficient, constraining the magnitude of inaccuracy in the computed intervalley scattering rate. Together, these observations indicate that the diffusion coefficient discrepancies may be attributed to factors which are not included in the ab-initio formulation of charge transport, for instance real-space gradients and space charge effects. This finding has relevance to the interpretation of diffusion coefficient measurements in terms of microscopic charge transport processes.

\section{Theory and Numerical Methods}

Our approach to solve for the high-field transport and noise properties of charge carriers has been described previously. \cite{Choi_2021,Cheng_2022,Hatanpaa_2023,Catherall_2023}
In brief, for a spatially homogeneous, non-degenerate electron gas subject to an applied electric field, the Boltzmann equation is given by

\begin{equation}
    \frac{q\mathbf{E}}{\hbar} \cdot \nabla_{\mathbf{k}} f_{\mathbf{k}} = - \sum_{\mathbf{k}'} \Theta_{\mathbf{k}\mathbf{k}'} \Delta f_{\mathbf{k}'} 
\end{equation}

Here, $q$ is the carrier charge, $\mathbf{E}$ is the electric field vector, $f_{\mathbf{k}}$ is the distribution function describing the occupancy of the electronic state indexed by wavevector $\mathbf{k}$, $\Delta f_{\mathbf{k}'} $ is the perturbation to the equilibrium electron distribution function $f_{\mathbf{k}}^{0}$, and $\Theta_{\mathbf{k}\mathbf{k}'}$ is the linearized collision matrix given by Eq. 3 of Ref.~\cite{Choi_2021}. We assume that only one band contributes to charge transport and thus neglect the band index.

For sufficiently large electric fields, the reciprocal space derivative of the total distribution function must be evaluated numerically. In the present formulation, the derivative is computed using a finite difference approximation given in Refs.~\cite{Mostofi_2008, Marzari_2012}. The BTE then takes the form of a linear system of equations (Eqn.~5 in Ref.~\cite{Choi_2021}) that can be solved by numerical linear algebra:

\begin{equation}
    \sum_{\mathbf{k}'} \Lambda_{\mathbf{k}\mathbf{k}'} \Delta f_{\mathbf{k}'} = \sum_{\gamma} \frac{qE_{\gamma}}{k_{B}T} v_{\mathbf{k},\gamma} f_{\mathbf{k}}^{0} 
\end{equation}

Here, $E_{\gamma}$ and $v_{\mathbf{k},\gamma}$ are the electric field and electron drift velocity along the $\gamma$-Cartesian axis, and the relaxation operator $\Lambda_{\mathbf{k}\mathbf{k}'}$ is defined as 

\begin{equation}
    \Lambda_{\mathbf{k}\mathbf{k}'} =  \Theta_{\mathbf{k}\mathbf{k}'} + \sum_{\gamma} \frac{qE_{\gamma}}{\hbar} D_{\mathbf{k}\mathbf{k}',\gamma}
\end{equation}

where $D_{\mathbf{k}\mathbf{k}',\gamma}$ is the momentum-space derivative represented in the finite-difference matrix representation given by Eqn.~24 in Ref.~\cite{Mostofi_2008}. The BTE is solved using numerical linear algebra to obtain the steady-state electron distribution function, from which transport properties can be obtained using an appropriate Brillouin zone sum. For instance, the drift velocity in the $\beta$ direction is given by

\begin{equation}
    V_{\beta} = \frac{1}{N} \sum_{\mathbf{k}} v_{\mathbf{k},\beta} f_{\mathbf{k}}
\end{equation}

where $N = \sum_{\mathbf{k}} f_{\mathbf{k}}$ is the number of electrons in the Brillouin zone.

The fluctuations in the occupancy of electronic states manifest in experiment as current noise, which can be characterized by the power spectral density (PSD). As derived in Ref.~\cite{Choi_2021}, the current PSD $S_{j_{\alpha}j_{\beta}}$ can be calculated at a given angular frequency $\omega$ as

\begin{equation} \label{eq:Sj}
 S_{j_{\alpha}j_{\beta}}(\omega) = 2 \bigg( \frac{2 e}{\mathcal{V}_0}\bigg)^2 \Re \bigg[\sum_{\kwv} v_{\kwv,\alpha} \sum_{\kwv'} (i\omega \mathbb{I} + \Lambda)^{-1}_{\kwv \kwv'} \bigg(f_{\mathbf{k}'}^s (v_{\kwv',\,\beta} - V_{\beta})\bigg)\bigg]
\end{equation}

where $j_{\alpha}$ and $j_{\beta}$ are the currents along axes $\alpha$ and $\beta$, $\mathcal{V}_0$ is the supercell volume, and $\mathbb{I}$ is the identity matrix. In the limit $\omega \tau \ll 1$, where $\tau$ is a characteristic relaxation time, $S_{j_{\alpha}j_{\beta}}$ is proportional to the diffusion coefficient, a relation known as the fluctuation-diffusion relation \cite{GGK_1979}. Therefore, the diffusion coefficient may also be computed from \cref{eq:Sj}.

In this work, we also computed the piezoresistivity of electrons in Si. For the calculations with compressive stress in the [001] direction, a small uniaxial compressive strain was applied in the [001] direction, and the other two lattice vectors were then relaxed. For the calculations with compressive stress in the [011] direction, the lattice vectors were changed manually until the desired stress state was reached.

The numerical details are as follows. The electronic structure and electron-phonon matrix elements are computed on a coarse $14 \times 14 \times 14$ grid using DFT and DFPT with \textsc{Quantum Espresso} \cite{Giannozzi_2009}. A finer coarse grid compared to the usual $8 \times 8 \times 8$ \cite{Hatanpaa_2023} was found to be necessary to converge the piezoresistivity. A wave-function energy cutoff of 40 Ryd was used for all calculations, and a relaxed lattice parameter of 5.431 $\textrm{\AA}$ was used for the unstrained properties. The electronic structure and electron-phonon matrix elements were interpolated onto the fine grid using \textsc{Perturbo} \cite{Zhou_2021}.

For temperatures of 160 -- 300 K, a grid density of $100 \times 100 \times 100$ for the electron states was used, while a grid density of $50 \times 50 \times 50$ was used for the phonons. Using a phonon grid with the same density as the electron grid resulted in mobility changes of 5\%, and using a grid density of $140 \times 140 \times 140$ for the electron states and $70 \times 70 \times 70$ for the phonon states resulted in mobility changes of 5\%. At these higher temperatures, an energy window extending up to 284 meV above the conduction band minimum was used with a Gaussian smearing parameter of 5 meV. Increasing the energy window to 342 meV resulted in a mobility change of 0.04\%. At 77 K, high-field PSD calculations were found to converge at a grid density of $140 \times 140 \times 140$ for electron states and $70 \times 70 \times 70$ for phonons, with a Gaussian smearing parameter of 2.5 meV and an energy window of 284 meV. However, for the piezoresistivity calculations in the low-field limit, a grid density of $500 \times 500 \times 500$ ($250 \times 250 \times 250$) for electrons (phonons) was required. In this case, an energy window of 20 meV was employed for computational tractability. The final linear system of equations used to obtain the mobility and the PSD was then solved by a Python implementation of the GMRES method \cite{Fraysse_2005}. For all calculations and temperatures, the Fermi level was adjusted to yield a carrier density of $4 \times 10^{13} \ \text{cm}^{-3}$. Spin-orbit coupling was neglected, as it has a weak effect on electron transport properties in Si \cite{Ma_2018, Ponce_2018}. Similarly, quadrupole electron-phonon interactions were neglected \cite{Park_2020}. For all calculations of the diffusion coefficient, a frequency of 1 GHz was used, selected so to ensure that $\omega \tau^{-1} \ll 1$, where $\tau$ is a characteristic relaxation time, while avoiding too low frequencies which result in numerical instabilities.


In our past work \cite{Hatanpaa_2023}, it has been shown that two-phonon scattering (2ph) is non-negligible in n-Si. Thus, for all PSD calculations, two-phonon scattering was included. For the piezoresistivity calculations, 2ph scattering could not be included due to the computational cost. However, we do not expect the absence of 2ph scattering for piezoresistivity to affect our conclusions, as it was shown in Ref.~\cite{Hatanpaa_2023} that the energy dependence of 2ph scattering rates exhibited the same qualitative trends as those of one-phonon rates, and further that most of the effect of 2ph scattering can be accounted for by scaling the 1ph scattering rates. As this scaling would be present at all applied stresses, we therefore do not expect that neglecting 2ph would affect the piezoresistivity values and our conclusions.

\section{Results}

\subsection{Electric-field dependence of hot electron diffusion coefficient}

We begin by examining the dependence of the diffusion coefficient on electric field at various temperatures. We first compare the experimental low-field values of the diffusion coefficient to the computed ones. We considered four temperatures (300, 200, 160, and 77 K), corresponding to those for which experimental data is available. At these temperatures, the computed (experimental) diffusion coefficients were 29.7 (37) $\text{cm}^2 \text{s}^{-1}$, 59.3 (62) $\text{cm}^2 \text{s}^{-1}$, 58.8 (71) $\text{cm}^2 \text{s}^{-1}$, and 1120 (141) $\text{cm}^2 \text{s}^{-1}$. For all temperatures besides 77 K, the computation underestimates the experimental data. The magnitude of the underestimate for $T>77$ K is consistent with a prior calculation of the electron mobility of Si when two-phonon scattering is included \cite{Hatanpaa_2023}. However, at 77 K, the computed value is $\sim 8 \times$ larger than experiment. This overestimate is possibly attributable to ionized impurity scattering which is neglected in the present calculations.

\begin{figure}[h]
\includegraphics[]{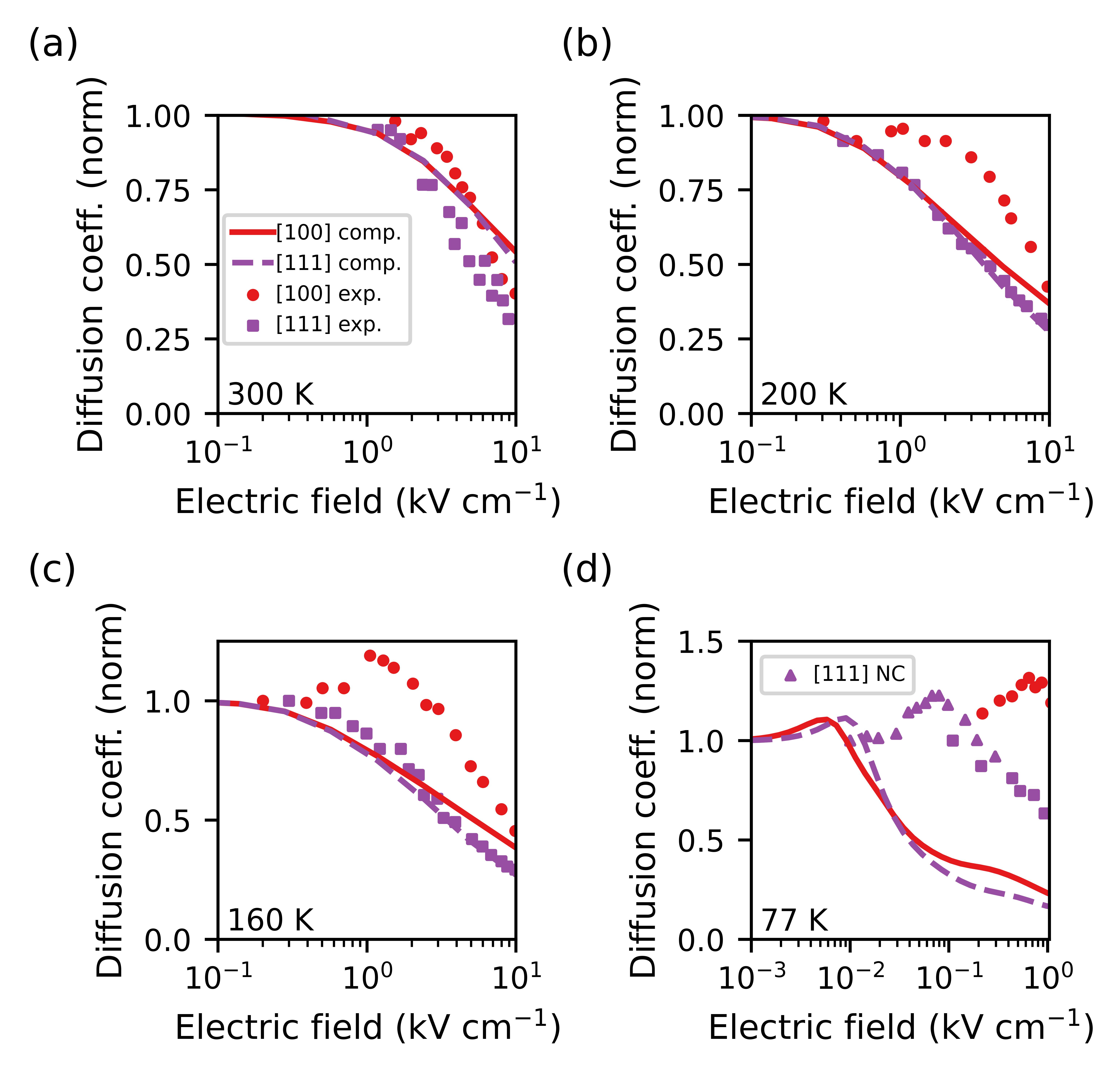}
{\phantomsubcaption\label{psd_vs_e_300}
\phantomsubcaption\label{psd_vs_e_200}
\phantomsubcaption\label{psd_vs_e_160}
\phantomsubcaption\label{psd_vs_e_77}}
\caption{ \label{psd} Normalized diffusion coefficient versus electric field at (a) 300 K, (b) 200 K, (c) 160 K, and (d) 77 K, with field applied along the [100] direction (red solid line) and [111] direction (purple dotted line). Experimental data along the [100] direction (red circles) and [111] direction (purple squares) from Figs.~3 and 4, Ref.~\cite{Brunetti_1981}. In (d), noise conductivity (NC) measurements (purple triangles) included for comparison at low electric fields.}
\label{psd_vs_e}
\end{figure}

To facilitate the comparison of trends with electric field in the subsequent plots, the computed data has been normalized to the calculated low-field diffusion coefficient, while the experimental data has been normalized to the value at the lowest electric field  reported. We note that due to the requirement that the transit time in the time-of-flight experiment be less than the dielectric relaxation time, no data was reported below a minimum field at each temperature \cite{Brunetti_1981}. The electric field dependence of the diffusion coefficient at 300 K is shown in \cref{psd_vs_e_300}. In experiment, it is observed that at low fields, the diffusion coefficient along the [100] and [111] directions are equal. Starting at less than 2 \kv, the diffusion coefficient along the [111] direction is less than in the [100], an anisotropy that has been attributed to intervalley diffusion \cite{Brunetti_1981}. The magnitude of this anisotropy continues to increase with field, reaches a maximum, and then decreases with field. The same qualitative trend with field is seen at 200 K in \cref{psd_vs_e_200}, with the main difference being the anisotropy manifesting at lower fields than at higher temperatures. At 160 K, shown in \cref{psd_vs_e_160}, there is a slight peak of the diffusion coefficient in the [100] direction at low fields and then a monotonic decrease for higher fields. At 77 K, shown in \cref{psd_vs_e_77}, initial increases of the diffusion coefficient with field are seen for both directions. 

The calculated results generally predict these trends qualitatively. At 300 K and 200 K, the correct trend of the anisotropy is reproduced, as the [111] diffusion coefficient is less than the [100] value once field values exceed 5 \kv~and 2 \kv, respectively. While the initial increase seen in experiment at 160 K with field applied in the [100] direction is not captured by computation, the qualitative anisotropy at high fields is reproduced. Similarly, at 77 K, the [111] diffusion coefficient is less than in the [100] once the electric field exceeds 0.2 \kv.

However, a number of quantitative discrepancies can be seen. At 160, 200, and 300 K, the computed anisotropy starts to manifest at higher fields than in experiment. In experiment, at 300 K the anisotropy is observed once the electric field exceeds 2 \kv, while at 200 K and 160 K the anisotropy manifests even below 1 \kv. Similarly, the magnitude of the anisotropy is underestimated, particularly for 160 K and 200 K, where the agreement with the [111] data is excellent, but the [100] data lies much above the computed values.

At 77 K, the qualitative behavior of the diffusion coefficient with field changes greatly. We note that the electric field range used in this calculation is smaller than in the other cases due to lack of convergence at high fields. In \cref{psd_vs_e_77}, for both directions measured an initial increase in the experimental PSD is seen. This increase is observed in computation, but at lower fields than in experiment. Given the relative importance of ionized impurity scattering at 77 K compared to higher temperatures, we examined whether the omission of this scattering mechanism in the calculation could play a role in the discrepancy. We implemented a simple model of ionized impurity scattering \cite{Long_1959} with a density of $10^{14}$ $\text{cm}^{-3}$. The non-monotonic features were observed to shift to higher fields, suggesting that ionized impurity scattering could be partly responsible for this discrepancy.


The anisotropy in the diffusion coefficient seen in experiment has been attributed to a mechanism known as intervalley diffusion. \cite{Brunetti_1981}
To understand this mechanism, consider the general expression for the intervalley diffusion coefficient $D^{\text{int}}$, given by  $D^{\text{int}} = n_1 n_2 (v_1 - v_2)^2  \tau_{\text{int}}$.  \cite{Price_1960,Brunetti_1981} Here,  $n_1$ and $n_2$ are the fractions of electrons in valleys of type 1 and 2, $v_1$ and $v_2$ are the drift velocities in valleys of type 1 and 2, and $\tau_i$ is the characteristic intervalley relaxation time. When the field is applied in the [111], the average velocities in each valley type are equal, and this extra contribution vanishes. While many transport properties such as mobility are insensitive to the balance between g- (between equivalent valleys) and f-type (between inequivalent valleys) scattering, $\tau_{\text{int}}$ is inversely proportional to the square of the f-type coupling constant \cite{Brunetti_1981}. A possible origin of the underpredicted anisotropy in the diffusion coefficient is therefore computed f-type scattering rates which are too large compared to experiment. To test this hypothesis, we compute other transport and noise properties which are sensitive to the distinct types of intervalley scattering.

\subsection{Microwave-frequency PSD}

\begin{figure}[h]
\includegraphics[]{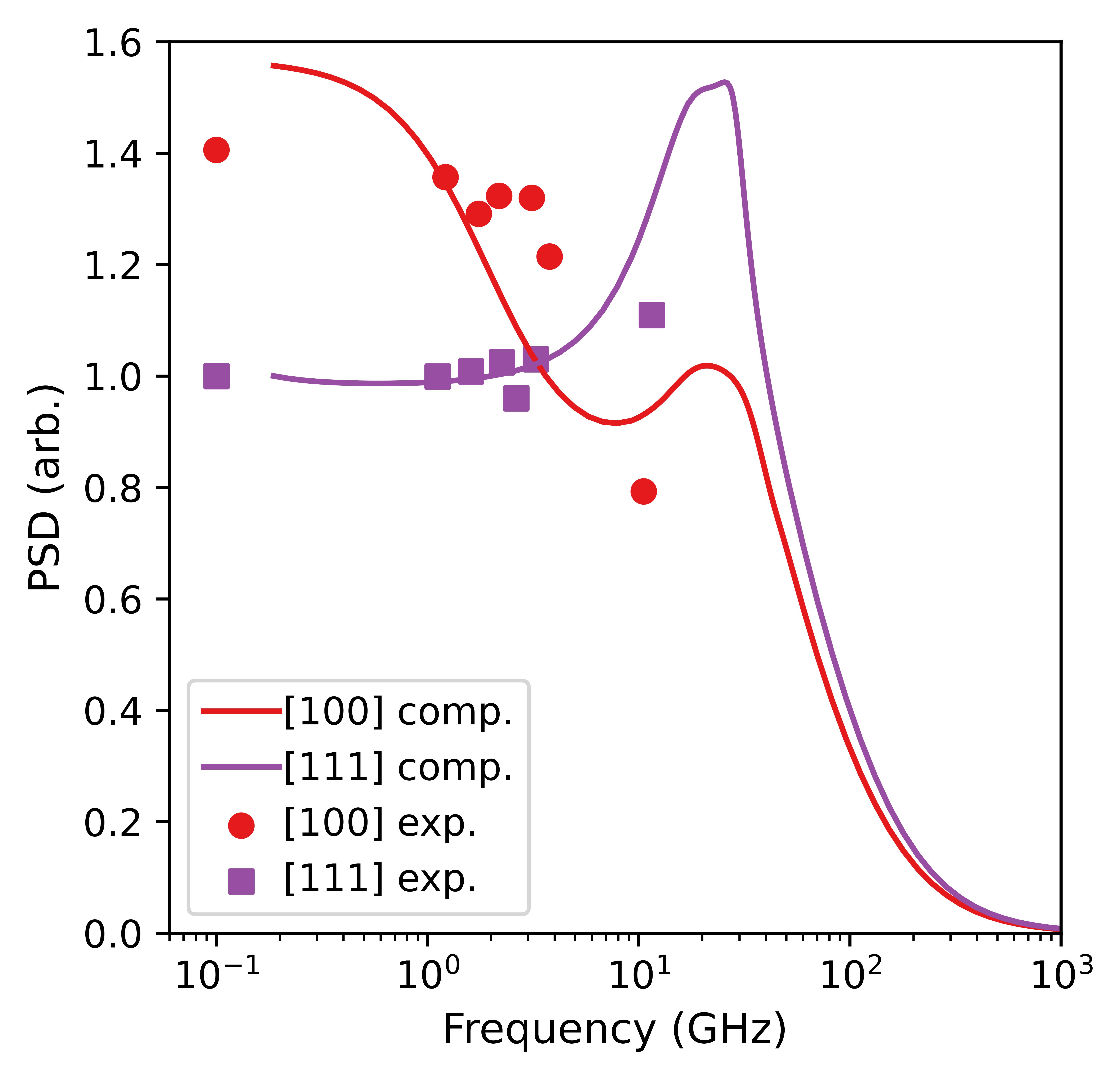}
\caption{Microwave PSD versus frequency at 77 K and 200 \vcm~applied electric field, with field applied along the [100] direction (red solid line) and [111] direction (purple solid line). Experimental data along the [100] direction (red circles) and [111] direction (purple squares) from Fig.~1, Ref.~\cite{Bareikis_1982}. In both cases, the data is normalized to the value of the PSD at the lowest frequency data point (computation, 0.19 GHz; experiment, 0.1 GHz) in the [111] direction.}
\label{psd_vs_f}
\end{figure}

We first compute the microwave-frequency ($\sim$0.1-100 GHz) PSD at 77 K and 200 \vcm, for which experimental data is available for comparison \cite{Bareikis_1982}. Here, the frequency ranges computed are much higher than those in which sources of noise such as 1/$f$ noise or generation-recombination noise would be relevant. However, if the frequency is comparable to an inverse time constant $\tau$ such as the momentum or energy relaxation time, non-monotonic features or roll-offs in the PSD with increasing frequency around frequencies satisfying $\omega \tau \sim 1$ will be observed \cite{Hartnagel_2001}. Comparing the frequencies at which these features occur therefore provides an independent test of the accuracy of the ab-initio diffusion coefficient calculations.

\Cref{psd_vs_f} shows the calculated spectral density of current fluctuations versus frequency, at 77 K and 200 \vcm~and with electric fields applied along the [111] and [100] directions, along with experimental data. At frequencies below 3 GHz, the [100] PSD is greater than the [111], due to the presence of intervalley noise. As intervalley scattering is characterized by a significantly smaller relaxation rate than either the energy or momentum relaxation rates, a rolloff in the [100] direction is observed around the relatively low frequency of 1 GHz. The presence of the ``convective" mechanism away from equilibrium rolls off at a frequency corresponding to the energy relaxation rate. Here, the convective peak occurs around 25 GHz. For semiconductors with a sublinear current-voltage characteristic, this convective contribution is negative \cite{Hartnagel_2001}. This mechanism is present in both the [100] and [111] cases, but is more obviously present in the [111] due to the lack of intervalley noise. Finally, as the frequency exceeds the momentum relaxation rate, the PSD rolls off to zero as the electronic system is not able to redistribute in response to the oscillating external field.  

Over the entire calculated frequency range, the computed results qualitatively capture the trends seen in experiment. The anisotropy seen at low frequencies due to intervalley noise, the rolloff in the [100] direction starting around 1 GHz due to frequency exceeding the characteristic intervalley scattering rate, and the convective noise peaks are all reproduced. (At frequencies above 100 GHz, the PSD is higher in the [111] direction, simply due to the greater mobility in this direction at 200 \vcm.) Relaxation times for the various noise sources (thermal, convective, and intervalley) can be obtained by fitting the computed curves to Lorentzians parameterized by the various relaxation times, as seen in Eq. 9.5 in Ref.~\cite{Hartnagel_2001}. For the [111] direction, an energy relaxation time of 15 ps was calculated using Monte Carlo simulation, as well as a momentum relaxation time of 2 ps, while our computation yields an energy relaxation time of 9 ps, and a momentum relaxation time of 4 ps \cite{Hartnagel_2001}. For the [100] direction, Monte Carlo simulation reported an energy relaxation time of 5 ps \cite{Hartnagel_2001}, and an intervalley relaxation time of 50 ps \cite{Bareikis_1982}, while our computation yields an energy relaxation time of 10 ps, and an intervalley relaxation time of 79 ps. The magnitudes of the relaxation times and relative difference between the momentum and energy relaxation times are thus in qualitative agreement with prior works. However, data only exists up to intermediate frequencies (around 10 GHz), so it is difficult to draw quantitative conclusions, especially for the momentum relaxation time.

As the difference in the PSD at low frequency is due to intervalley noise, and the magnitude of this difference is captured accurately by our computation, the results of \cref{psd_vs_f}  suggest that the f-type scattering rates in computation are compatible with their actual values. In addition, the frequency of the intervalley roll-off and convective mechanism being well-captured imply that both the intervalley and energy relaxation rates are qualitatively consistent with experimental values. 

\subsection{Piezoresistivity}

\begin{figure}[h]
\includegraphics[]{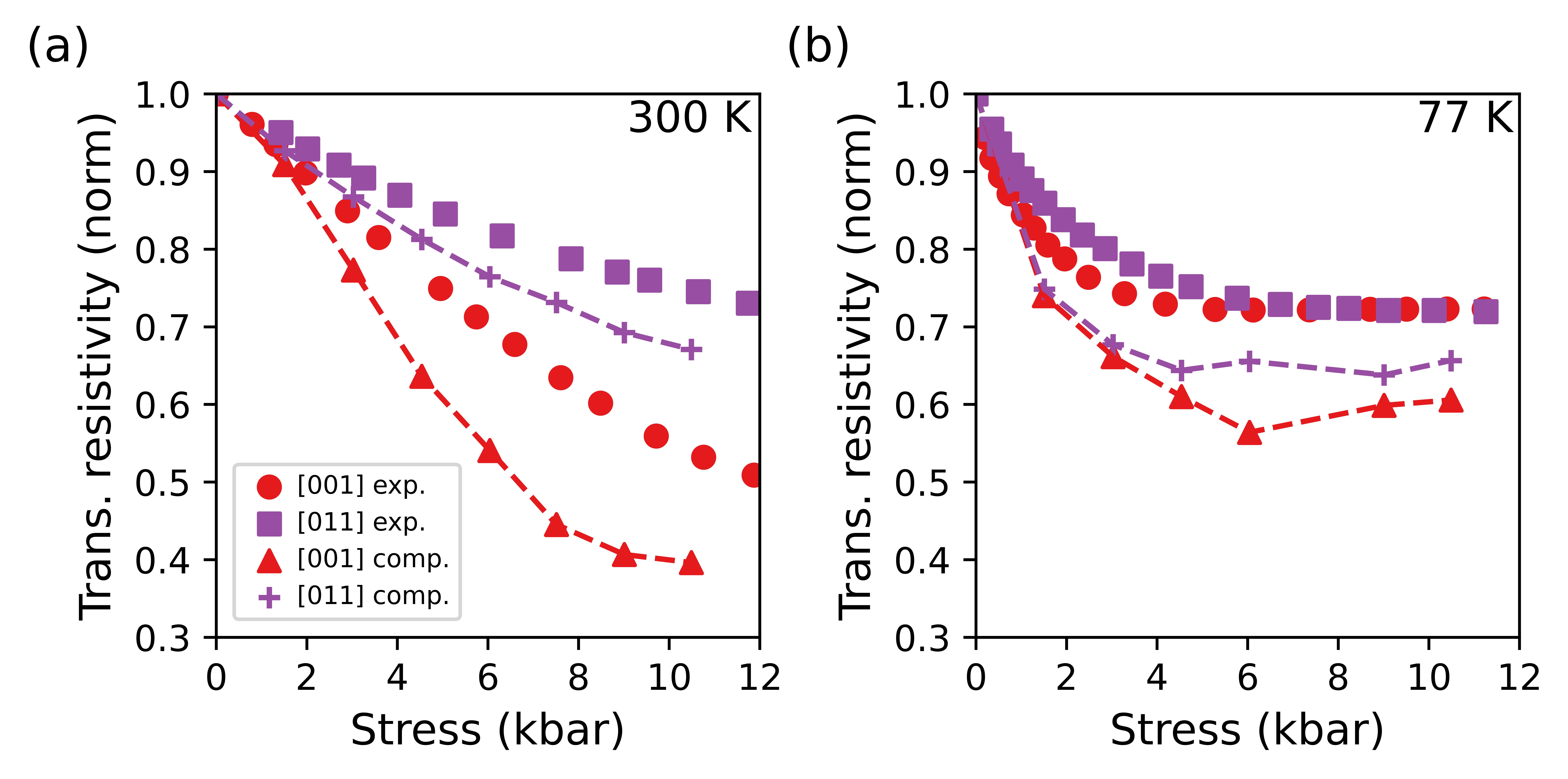}
{\phantomsubcaption\label{resistivity_300}
\phantomsubcaption\label{resistivity_77}}
\caption{\label{resist} Computed normalized transverse resistivity versus stress at (a) 300 K and (b) 77 K,  with stress applied along the [001] direction (red triangles) and [011] direction (purple crosses). Experimental data along the [001] direction (red circles) and [011] direction (purple squares) from Figs.~3 and 4, Ref.~\cite{Hansen_1974}.}
\label{resistivity}
\end{figure}

We next compute the piezoresistivity at 300 K and 77 K, for which experimental data is available \cite{Hansen_1974}. It has been shown that when sufficiently high stress is applied along the [001], f-type scattering is eliminated; while when high stress is applied along the [011], 50\% of the original f-type scattering (compared to the unstressed case) remains. This behavior arises from the energy shifts of the initially isoenergetic valleys with stress. If f-type scattering is negligible (such as at low temperatures), the transverse mobility (when the field is applied along the [100] direction) at high stress in both cases is expected to be identical \cite{Jorgensen_1978}, while in cases where f-type scattering is non-negligible such as room temperature, the resistivity in the case with stress applied along the [011] will be greater than in the [001]. \Cref{resistivity} shows experimental data consistent with these expectations, where at 300 K in \cref{resistivity_300} the high-stress resistivity is greater in the [011] case due to the presence of f-type scattering, while at 77 K in \cref{resistivity_77} the high-stress transverse resistivity along the two directions is identical, indicating negligible f-type scattering at this temperature.

In \cref{resistivity_300}, the computed transverse resistivity versus stress in the [001] and [011] directions at 300 K is presented. The computed anisotropy exhibits qualitative agreement with experiment, as in both cases the resistivity is less when the stress is applied along the [001] compared to the [011] case. Due to the non-negligible contribution of f-type scattering  at 300 K, applying pressure along the [001] decreases the magnitude of intervalley scattering by a greater amount than in the [011] case, leading to a greater [011] transverse mobility (and thus lower resistivity) than when stress is applied along the [001]. However, the computation underpredicts the resistivity at all pressures for both applied stress directions.

In \cref{resistivity_77}, the computed transverse resistivity versus stress at 77 K is shown along with experimental data. Here, it is observed in experiment that at high stresses, the resistivity along both directions saturates to closer to the same value than at 300 K. The computed resistivity saturates with pressure to a slightly lower value than in experiment, but the difference between the two directions is considerably smaller than at 300 K (69\% at 300 K versus 8\% at 77 K).

The relatively small difference in the high-pressure 77 K resistivity between the two directions indicates that f-type scattering is negligible at this temperature, while at 300 K the computed difference between the two directions is comparable with experimental results. The agreement at both temperatures indicates that the magnitude of f-type scattering at these temperatures is being qualitatively captured.

\section{Discussion}

We now discuss our findings in the context of intervalley scattering and noise measurements in semiconductors. \Cref{psd_vs_e} indicates that the  anisotropy of the diffusion coefficient in n-Si is qualitatively captured in the calculation, with the diffusion coefficient in the [111] direction being less than in the [100] in the high-field limit for all temperatures measured. The primary discrepancies between experiment and computation between 160 and 300 K are the anisotropy in the computed results being smaller and not manifesting until higher fields than in experiment. The smaller anisotropy in the computed results suggests that the computation underestimates the amount of intervalley noise, and thus overestimates the amount of f-type scattering. However, \cref{psd_vs_f} indicates that the computed intervalley scattering rates and intervalley noise magnitude are qualitatively compatible with experiment, and \cref{resistivity} shows that the variation of f-type scattering with temperature is qualitatively captured as well. The amount of error in the computed f-type scattering rate is therefore constrained to values that are insufficient to explain the discrepancies in the diffusion coefficient.



We suggest that the discrepancies may arise from the neglect of spatial inhomogeneities present in experiment. The ab-initio method usually does not include real-space effects such as concentration gradients or space charge effects. Although the time-of-flight experiment was carefully implemented to avoid dielectric relaxation in the sample, it is conceivable that fluctuations in drift velocity associated with intervalley scattering within the generated electron pulse could lead to space charge effects which would spatially broaden the pulse and hence increase the measured diffusion coefficient. This effect would be present only in the [100] direction due to the absence of intervalley scattering in the [111] direction. Further, these effects would not appear in the microwave PSD as these frequencies are much higher than those associated with any dielectric relaxation phenomena. Additional study will be required to determine the origin of the diffusion coefficient discrepancies.



\noindent
\section{Summary} 

We have computed the hot-electron diffusion coefficient, microwave PSD, and piezoresistivity in Si from first principles from 77 -- 300 K. We find that while qualitative features of the diffusion coefficient such as the anisotropy at high electric fields are generally predicted, several trends of the calculated values differ from experiment. We computed the piezoresistivity and microwave PSD to investigate whether an inaccurate description of f-type intervalley scattering could explain the discrepancies. However, the good qualitative agreement of these properties with experiment excluded this possibility, leading to the hypothesis that the measured diffusion coefficient is influenced by factors not included in ab-initio calculations such as real-space gradients and space charge effects. This finding indicates that care must be taken when interpreting diffusion coefficient measurements in terms of microscopic charge transport processes.

\begin{acknowledgements}
B.H. was supported by a NASA Space Technology Graduate Research Opportunity. A.J.M. was supported by AFOSR under Grant Number FA9550-19-1-0321. The authors thank J.~Sun, S.~Sun, D.~Catherall, and T.~Esho for helpful discussions.
\end{acknowledgements}

\bibliography{references}

\end{document}